\documentclass[pra,aps,twocolumn,epsfig,superscriptaddress,showpacs]{revtex4}
\usepackage{amsmath}
\usepackage{amssymb}
\usepackage[dvips]{graphicx}
\usepackage{dcolumn}
\usepackage{bm}
\usepackage[colorlinks=false,dvipdfm]{hyperref} 
\usepackage{setspace}
\usepackage{amsfonts}
\usepackage{rotating}
\usepackage{makeidx}
\usepackage{amsthm}
\usepackage{CJK}

\newcommand{\tr}{\mbox{Tr}}
\newcommand{\Exp}{\mbox{Exp}}

\begin{document}

\title{Irreducible Multi-Qutrit Correlations in Greenberger-Horne-Zeilinger Type States \footnote{Physical Review A, 2011, 84(6):062328.
\\http://link.aps.org/doi/10.1103/PhysRevA.84.062328}}

\author{Fu-Lin Zhang }\email[Email: ]{flzhang@tju.edu.cn}
\affiliation{Physics Department, School of Science, Tianjin
University, Tianjin 300072, China}

\author{Jing-Ling Chen }
\email[Email: ]{chenjl@nankai.edu.cn}\affiliation{Theoretical
Physics Division, Chern Institute of Mathematics, Nankai University,
Tianjin, 300071, China} \affiliation{Centre for Quantum
Technologies, National University of Singapore, 3 Science Drive 2,
Singapore 117543}

\date{\today}

\begin{abstract}


Following the idea of the continuity approach in [D. L. Zhou, Phys.
Rev. Lett. 101, 180505 (2008)], we obtain the degrees of irreducible
multi-party correlations in two families of $n$-qutrit
Greenberger-Horne-Zeilinger type states. For the pure states in one
of the families, the irreducible $2$-party, $n$-party and
$(n-m)$-party ($0< m < n-2$) correlations are nonzero, which is
different from the $n$-qubit case. We also derive the correlation
distributions in the $n$-qutrit maximal slice state, which can be
uniquely determined by its $(n-1)$-qutrit reduced density matrices
among pure states. It is proved that there is no irreducible
$n$-qutrit correlation in the maximal slice state. This enlightens
us to give a discussion about how to characterize the pure states
with irreducible $n$-party correlation in arbitrarily
high-dimensional systems  by the way of the continuity approach.
\end{abstract}



\maketitle




\section{Introduction}

Coherent superposition is the essential distinction between a
quantum system and a classical one. This distinction becomes more
significant in composite systems, which appears in the non-classical
correlations in the quantum systems. Many concepts have been
developed to describe these correlations, such as the entanglement
\cite{EPR} which depicts the nonseparability of the state of a
composite quantum system. Another concept refers to the nonlocality
is characterized by violation of a Bell inequality \cite{Bell},
which means that the local measurement outcomes of the state cannot
be described by a local hidden variables model.

In the information-based viewpoint, the correlation in a quantum
system can be viewed as the relationship of the whole system and its
subsystems. Namely, it measures the degree a quantum state can be
described by the reduced states of its subsystems. The total
correlation \cite{Correlation} in a multipartite quantum system has
been defined as the difference between the sum of the von Neumann
entropies of all the subsystems and that of the whole system, while
the so-called quantum discord
\cite{PhysRevLett.88.017901,henderson2001classical,datta2008quantum},
widely studied in very recent years, was considered to be the
quantum part (opposite to the classical one) of the total
correlation.
In the current paper, we concerns us in another alternative
classification, in which the total correlation in a multiparty
system is divided into different levels, namely pairwise, triplewise
and so forth.

Linden \emph{et al.} \cite{Linden02} proposed the concept of
irreducible $n$-party correlation in an $n$-partite quantum state.
This concept is based on the principle of maximum entropy and
describes how much more information in the $n$-party level than what
is contained in the $(n-1)$-partite reduced states. A surprising
result was given in the original work of Linden and his
collaborators \cite{Linden02,Linden02W} that almost all $n$-party
pure states are determined by their reduced density matrices. In
$n$-qubit case, the only pure states that can't be determined by
their reduced density matrices are proved to be the generalized
Greenber-Horne-Zeilinger (GHZ) states \cite{WL1,WL2}. This indicates
that among $n$-qubit pure states, only in the case of the
generalized GHZ states the irreducible $n$-party correlation has a
nonzero value \cite{Zhou08}. For the arbitrarily high-dimensional
system, Feng \emph{et al.} \cite{QIC} introduced the generalized
Schmidt decomposition (GSD) states and proved them to be the
$n$-partite pure states undetermined \emph{among pure states} by
their reduced density matrices. It still remains an open question
whether the GSD states identified in \cite{QIC} are precisely the
pure states undetermined by their reduced density matrices
\emph{among arbitrary states} (pure or mixed). In other words, it is
under confirmation that, whether the irreducible $n$-party
correlation in a $n$-partite non-GSD states is nonzero or not.

In Zhou's recent work \cite{Zhou08}, the concept of irreducible
$n$-party correlation has been generalized to $m$-party ($2 \leq
m\leq n$) levels, where a classification of the total correlation in
an $n$-partite state is constructed. For a given $n$-partite quantum
state $\rho^{[n]}$, Zhou introduced a sequence of density matrices.
The $m$-th one, $\tilde{\rho}^{[n]}_m$ ($1 \leq m\leq n$), has the
same $m$-party reduced density matrix as $\rho^{[n]}$, and the
maximal value of von Neumann entropy, which is considered to contain
the $m$-party level information of the given state without the
higher level information. The degree of irreducible $m$-party
correlation is defined as
\begin{eqnarray}\label{mcorrlation}
C^{(m)}(\rho^{[n]})=S(\tilde{\rho}^{[n]}_{m-1})-S(\tilde{\rho}^{[n]}_m),
\end{eqnarray}
where $S(\sigma)=-\tr( \sigma \ln \sigma )$ is the von Neumann
entropy.
For the states $\rho^{[n]}$ with
maximal rank, $\tilde{\rho}^{[n]}_m$ is proved to have a
\emph{standard exponential form}
\begin{eqnarray}\label{exp}
\tilde{\rho}^{[n]}_m=\Exp \bigr(Q^{[m]}\bigr),
\end{eqnarray}
where $ Q^{[m]}$ is a sum of $m$-partite hermitian operators.
Directly, $\tilde{\rho}^{[n]}_n=\rho^{[n]}$ is itself and
$\tilde{\rho}^{[n]}_1 = \bigotimes ^{n}_{i=1} \rho^{(i)}$ is the
direct product of all the single partite reduced density matrices.
To derive the degrees of the irreducible multiparty correlations in
quantum states with nonmaximal ranks, Zhou presented a continuity
approach based on the fact that a multipartite state without maximal
rank can always be regarded as the limit of a series states with
maximal rank, such $\rho^{[n]}(\gamma) |_{\gamma \rightarrow
+\infty}=\rho^{[n]}$. By constructing the sequence of density
matrices $\tilde{\rho}^{[n]}_m (\gamma)$, one can obtain the degree
of irreducible $m$-party correlation of $\rho^{[n]}(\gamma)$
\begin{eqnarray}\label{mcorrlation-gamma}
C^{(m)}(\rho^{[n]}(\gamma))=S(\tilde{\rho}^{[n]}_{m-1}(\gamma))-S(\tilde{\rho}^{[n]}_m(\gamma)).
\end{eqnarray}
Then $C^{(m)}(\rho^{[n]})=C^{(m)}(\rho^{[n]}(\gamma ))|_{\gamma
\rightarrow +\infty}$ give the correlations of the state
$\rho^{[n]}$.

In this approach, Zhou gave the correlation distributions of the
$n$-qubit stabilizer states and the generalized GHZ states. It is
worth noting that, in the $n$-qubit generalized GHZ states, which
are the only pure states with nonzero $n$-qubit correlation
\cite{WL1,WL2}, only irreducible $2$-party and $n$-party
correlations have nonzero values. However, a systematic method to
construct the standard exponential form density matrices in Eq.
(\ref{exp}) for a given state with maximal rank has not been found.
Consequently it is difficult to analytically obtain the correlation
distributions in the states without maximal ranks.
To the best of our knowledge, the known correlation distributions,
both the analytical \cite{Zhou08} and the numerical
\cite{zhou2009efficient}, in multipartite quantum states are
restricted in $n$-qubit systems. The main purpose of this paper is
to derive the degrees of irreducible multiparty correlations of some
typical quantum states without the maximal rank in $n$-qutrit
system. On one hand, through the analysis of some special examples,
the difference between the correlation distributions in qubit
systems and the ones in high-dimensional systems can be revealed. On
the other hand, more analytical results could contribute to the
construction a systematic method to calculate the degrees of
irreducible multiparty correlations, at least for a family of
states.

Our main results are given in the second section. As the first
trial, we concerns us in two families of $n$-qutrit GHZ-like sates,
in which the pure states belongs to the GSD states defined in
\cite{QIC}. The maximal slice (MS) states \cite{MS} can be viewed as
another generalization of the original GHZ states. We introduce an
$n$-qutrit version MS state, which is non-GSD according to the
results in \cite{QIC}, and obtain its multiparty correlation
distributions. Based on these results, in the last section we give a
discussion about the feasibility to solve the open problem about the
GSD states in the continuity approach.

\section{Multi-Qutrit Correlations}


To derive the degrees of irreducible multiparty correlations in the
three families of $n$-qutrit states studied in this paper, we adopt
Zhou's continuity approach with a little improvement. Namely, the
results in the original work of Zhou \cite{Zhou08} indicated that,
the standard exponential form state $\tilde{\rho}^{[n]}_m(\gamma)$
contains the maximal von Neumann entropy among the states with the
same $m$-party reduced density matrices, and this property is holden
when its parameter $\gamma \rightarrow +\infty$. Therefore, for a
given state, $\rho^{[n]}$, with nonmaximal rank, in stead of
constructing the series states $\rho^{[n]}(\gamma)$ and
corresponding $\tilde{\rho}^{[n]}_m(\gamma)$,  we construct a
sequence of states $\sigma^{[n]}_{m}(\gamma_m)$ in the standard
exponential form (\ref{exp}) whose limit
$\sigma^{[n]}_{m}|_{\gamma_m \rightarrow +\infty}$ has the same
$m$-party reduced density matrices as $\rho^{[n]}$. Here, for
different $m$, the states $\sigma^{[n]}_{m}(\gamma_m)$ are
independent, and the same are true for their parameters $\gamma_m$.
Then the degree of the irreducible $m$-party correlation in the
state $\rho^{[n]}$ is given by
\begin{eqnarray}\label{mcorrlation-sigma}
C^{(m)}(\rho^{[n]})=S(\sigma^{[n]}_{m-1}|_{\gamma_{m-1} \rightarrow
+\infty})-S(\sigma^{[n]}_{m}|_{\gamma_m \rightarrow +\infty}).
\end{eqnarray}

%

\subsection{First GHZ-type states}

 We introduce the $n$-qutrit states
in the subspace $\{|0^{[n]} \rangle, |1^{[n]} \rangle, |2^{[n]}
\rangle \}$ as the first family of GHZ-type states, where $|i^{[n]}
\rangle = |ii...i\rangle$ with $i=0,1,2$, denotes the direct product
of the basis $|i\rangle$ for $n$ qutrits. They can be expressed as
\begin{eqnarray}\label{G1}
\mathcal{G}=\sum_{i=0,j=0}^{2,2} c_{ij} |i^{[n]} \rangle \langle
j^{[n]}|,
\end{eqnarray}
with $c_{ij}=c_{ij}^{*}$ and the positive real numbers $c_{ii}$
satisfying $\sum c_{ii}=1$. One can write the diagonal elements in
spherical coordinate as $(c_{00},c_{11},c_{22})=(\sin^2 \theta
\cos^2 \phi,\cos^2 \theta,\sin^2 \theta \sin^2 \phi)$ with
$\theta,\phi \in [0,\pi/4]$.

\emph{Theorem 1.} The degrees of irreducible multiparty correlations
in the $n$-qutrit GHZ-type state $\mathcal{G}$ in Eq. (\ref{G1}) are
given by
\begin{eqnarray}\label{CG1}
C^{(2)}&=&(n-1) \mathcal{H}_{3} (\theta,\phi), \nonumber \\
C^{(n)}&=&\mathcal{H}_{3} (\theta,\phi)- S(\mathcal{G}),
\end{eqnarray}
and $C^{(m)}=0$ for $m=3,4,...n-1$. Here, $\mathcal{H}_{3}
(\theta,\phi)  =\mathcal{H}_2(\theta)+\sin^2 \theta
\mathcal{H}_2(\phi)$ denotes the trinary entropy of the
probabilities $\{\cos^2 \theta,\sin^2 \theta \cos^2 \phi,\sin^2
\theta \sin^2 \phi \}$, with $\mathcal{H}_2(\alpha)=-\cos^2 \alpha
\ln \cos^2 \alpha - \sin^2 \alpha \ln \sin^2 \alpha $ being the
binary entropy.

\begin{proof}
Let $Z_{j}=| 0 \rangle \langle 0 | - | 2 \rangle \langle 2 |$ be the
spin-$1$ operator in $z$-axis of the $j$-th qutrit, the $2$-partite
operators defined as
\begin{eqnarray}\label{Qij}
Q_{ij}=\frac{2}{3} \bigr[ \frac{1}{2} + \cos \frac{2 \pi}{3} (Z_i -
Z_j) \bigr]
\end{eqnarray}
satisfies $Q_{ij}^2=Q_{ij}$ and $\tr_{i,j}  Q_{ij} =3$. We construct
an $n$-qutrit state
\begin{eqnarray}\label{ExpG1}
\sigma_{g} (\gamma)= \Exp \bigr( \eta + \gamma \sum^{n}_{j=2}Q_{1j}
- \gamma_{1} Z_{1}^2 + \gamma_{2} Z_1 \bigr),
\end{eqnarray}
where $\tanh \gamma_{2} = \cos 2\phi $ and $\exp\gamma_1=2 \cosh
\gamma_{2} \cot^2 \theta$, and the value of $\eta$ is determined by
the normalization condition $\tr \sigma_{g}(\gamma) =1$.
Straightforward calculation gives
\begin{eqnarray}\label{ExpG11}
\sigma_{g} (\gamma)&=& \prod^{n}_{j=2}\bigr(\frac{1}{e^{\gamma}+2} +
\frac{e^{\gamma}-1}{e^{\gamma}+2} Q_{1j}  \bigr) f(Z_1),
\end{eqnarray}
with $f(Z_1)=  \cos^2 \theta (1-Z_1^2)+\frac{1}{2} \sin^2 \theta (
Z_1^2 + \cos 2 \phi Z_1 ) $. When $\gamma$ approaches infinity the
limit of $\sigma_{g} (\gamma)$ is nothing but the diagonal terms of
$\mathcal{G}$, i.e., $\sigma_{g}|_{\gamma \rightarrow +\infty }
=\mathcal{D}_{g}=\sum c_{ii} |i^{[n]} \rangle \langle i^{[n]}|$.

The state $\mathcal{D}_{g}$ has the same $(n-1)$-partite reduced
matrices as the states $\mathcal{G}$, and there exists only
irreducible $2$-party correlation in $\sigma_{g}(\gamma)$.
Therefore, one can take $\tilde{\mathcal{G}}_m = \mathcal{D}_{g}$
for $m=2,3,...n-1$ and obtain the results in Eq. (\ref{CG1}).
\end{proof}

The pure states in this family are always equivalent to the
generalized GHZ states for $n$-qutrit system
\begin{eqnarray}\label{G1p}
|\mathcal{G}^p\rangle = \cos \theta |0^{[n]} \rangle + \sin \theta
\cos \phi |1^{[n]} \rangle + \sin \theta \sin \phi |2^{[n]} \rangle
\end{eqnarray}
under local unitary transformations, which belongs to the GSD states
in \cite{QIC} apparently. There are $n \mathcal{H}_3 (\theta,\phi)$
correlations in $|\mathcal{G}^p\rangle$,
$\mathcal{H}_3(\theta,\phi)$ of which is irreducible $n$-party
correlation and the others belongs to the $2$-party level. This
distribution is the same as the generalized $n$-qubit GHZ states in
\cite{Zhou08}. When $\phi=0$,
$\mathcal{H}_3(\theta,0)=\mathcal{H}_{2}(\theta)$, this result
returns to the $n$-qubit case.

\subsection{Second GHZ-type States}

 A generalization of the family of
$n$-qutrit states $\mathcal{G}$ is the one in the subspace $\{
|0^{[n]}\rangle, |1^{[n]}\rangle ,|0^{[m]}2^{[n-m]}\rangle \}$ with
$m$ being a positive integer less than $n$, and
$|0^{[m]}2^{[n-m]}\rangle= |0^{[m]}\rangle \otimes
|2^{[n-m]}\rangle$ denoting the direct product of basis $|0\rangle$
for the first $m$ qutrits
 and $|2\rangle$ for the others. Denoting the basis $(|\bar{0}\rangle,|\bar{1}\rangle,|\bar{2}\rangle)=(|0^{[n]}\rangle, |1^{[n]}\rangle
,|0^{[m]}2^{[n-m]}\rangle)$, the states in this family can be
written as
\begin{eqnarray}\label{G2}
\mathcal{G}_{2}=\sum^{2,2}_{i=0,j=0} c_{ij} |\bar{i}\rangle \langle
\bar{j}|,
\end{eqnarray}
with the same constraint on $c_{ij}$ as Eq. ({\ref{G1}}), and the
diagonal elements $c_{ii}$ also can be expressed in the spherical
coordinate $\theta$ and $\phi$.

\emph{Theorem 2.} The degrees of irreducible multiparty correlations
in the second family of GHZ-type state $\mathcal{G}_2$ in Eq.
(\ref{G2}) are given by
\begin{eqnarray}\label{CG2}
C^{(2)}&=& m \mathcal{H}_{2}(\theta)+(n-m-1)\mathcal{H}_{3}(\theta,\phi), \nonumber\\
C^{(n-m)}&=& \mathcal{H}_{3}(\theta,\phi)-\mathcal{H}_{3}(\theta,\varphi), \nonumber\\
C^{(n)}&=& \mathcal{H}_{3}(\theta,\varphi)-S(\mathcal{G}_{2}),
\end{eqnarray}
and $C^{(k)}=0$ for the other integer numbers $2 \leq k \leq n$ ,
where the value of $\varphi$ is given by $ \cos^2 2 \varphi= \cos^2
2 \phi + 4 |c_{02}|^2 /\sin^4 \theta$.

\begin{proof}
The quantum states $\mathcal{D}_{2}=\sum c_{ii}|\bar{i}\rangle
\langle \bar{i}|$ has the same $k$-partite reduced matrices as
$\mathcal{G}_{2}$ for $k < n-m$. It can be proved to be the limit of
a state in the form (\ref{exp}).

Let us construct a $2$-patite operator $P_{ij}=(2Z^2_i-1)
\lambda^{(3)}_j$ by using the spin operator $Z_i$ and the third
Gell-Mann matrix of the $j$-th qutrit, $\lambda^{(3)}_j= |0\rangle
\langle 0 | - |1 \rangle \langle 1 |$. It satisfies
$P_{ij}^2={\lambda^{(3)}_j}^2$ and  $P_{ij}^3=P_{ij}$. Then, the
basis $|\bar{0}\rangle$, $|\bar{1}\rangle$ and $|\bar{2}\rangle$ are
the three eigenvectors of $\Omega=  \sum^{m}_{j=1} P_{m+1,j} +
 \sum^{n}_{l=m+2} Q_{m+1,l}$, corresponding to the maximal eigenvalue
 $\omega_{max}=n-1$. Choosing the values of $\gamma_1$ and
 $\gamma_2$ the same as the ones in Eq. (\ref{ExpG1}),
  and $\exp \eta=(2 \cosh \gamma+1 )^{-m} (\exp \gamma +2)^{-n+m+1} (\exp \gamma_1 + 2 \cosh \gamma_2)^{-1}$, the quantum state with
 only irreducible $2$-party correlation
\begin{eqnarray} \label{ExpG2}
\sigma_2 (\gamma)= \Exp \bigr(  \eta+ \gamma \Omega - \gamma_1
Z^2_{m+1} +\gamma_2 Z_{m+1} \bigr)
\end{eqnarray}
has the limit $\sigma_2|_{\gamma \rightarrow + \infty}
=\mathcal{D}_2$. Accordingly, we can choose
$\tilde{\mathcal{G}}_{2,k}=\mathcal{D}_2$ for $ k=2,3,...n-m-1$.

To obtain $\tilde{\mathcal{G}}_{2,k}$ for the other values of $k$,
we introduce three $(n-m)$-partite operators
$\Sigma_1=\prod^{n}_{j=m+1} \lambda^{(4)}_j$,
$\Sigma_2=\lambda^{(5)}_{m+1}\prod^{n}_{j=m+2} \lambda^{(4)}_j$ and
$\Sigma_3=Z_{m+1}\prod^{n}_{j=m+2} Z^2_j$, with $\lambda^{(4)}_j =
|0 \rangle \langle 2|+ |2 \rangle \langle 0|$ and $\lambda^{(5)}_j =
-i|0 \rangle \langle 2|+ i|2 \rangle \langle 0|$ being the fourth
and fifth Gell-Mann matrices of the $j$-th qutrit. They satisfy the
relations $\Sigma_\alpha \Sigma_\beta =i \epsilon_{\alpha \beta
\gamma} \Sigma_\gamma $, $\Sigma_{\alpha}^2=\prod^{n}_{j=m+1} Z^2_j$
and $[\Sigma_\alpha , \Omega ]=0$, where
$\alpha,\beta,\gamma=1,2,3$, and can be viewed as the Pauli matrices
in the subspace $\{ |\bar{0} \rangle, |\bar{2} \rangle \}$. Let
$\tanh \gamma_{2} = \cos 2\varphi $ and $\exp\gamma_1=2 \cosh
\gamma_{2} \cot^2 \theta$, we can define an $n$-qutrit density
matrix
\begin{eqnarray} \label{ExpG22}
\tau_{2} (\gamma)= \Exp \bigr(  \eta+ \gamma \Omega - \gamma_1
\Sigma^2_{r} +\gamma_2 \Sigma_{r} \bigr),
\end{eqnarray}
where $\Sigma_{r}=\cos\xi  \Sigma_{3} + \sin \xi  \cos \zeta
\Sigma_{1} + \sin \xi \sin \zeta \Sigma_{2} $ with the parameters
$\zeta=\frac{1}{2} \arg (c_{20}/c_{02})$ and $ \xi=\arccos(
\cos2\phi / \cos 2 \varphi)$, and $\eta$ is determined by the
normalization condition $\tr \tau_{2}(\gamma)=1 $. It has no
irreducible $(n-m+1)$-party or higher level correlation, and
approaches the quantum state $\mathcal{B}_2=\mathcal{D}_2+c_{02}
|\bar{0}\rangle \langle \bar{2}|+c_{20} | \bar{2}\rangle \langle
\bar{0} |$ when the parameter $\gamma \rightarrow + \infty$. The
$n$-qutrit state $\mathcal{G}_2$ has the same $(n-1)$-partite
reduced density matrices as $\mathcal{B}_2$. Therefore, we can take
$\tilde{\mathcal{G}}_{2,k}=\mathcal{B}_2$ for $ k=n-m,n-m+1,...n-1$.
One can yield the results in Eq. (\ref{CG2}) via a direct
calculation.
\end{proof}

For the pure state in this family
\begin{eqnarray}\label{G2p}
|\mathcal{G}^p_{2}\rangle =\sin \theta \cos \phi |\bar{0} \rangle +
\cos \theta|\bar{1} \rangle + \sin \theta \sin \phi |\bar{2}
\rangle,
\end{eqnarray}
the variable $\varphi=0$, the total correlation with the value
$C^{T}=m \mathcal{H}_2 (\theta)+(n-m)\mathcal{H}_{3}(\theta,\phi)$
is divided into three nonzero irreducible multi-qutrit correlations
as $C^{(2)}=m \mathcal{H}_2
(\theta)+(n-m-1)\mathcal{H}_{3}(\theta,\phi)$, $C^{(n-m)}= \sin^2
\theta \mathcal{H}_2 (\phi) $ and $C^{(n)}=\mathcal{H}_{2}(\theta)$.
This result is different with the $n$-qubit case, which indicates
there could exist nonzero $C^{(k)} (2<k<n)$ in an $n$-qutrit pure
states with the irreducible $n$-party correlation.

When $\phi=0$, the state (\ref{G2p}) becomes the $n$-qubit
generalized GHZ state, and at the same time the above correlation
distribution returns the corresponding one. When $\theta=\pi/2$, the
state (\ref{G2p}) equivalents to the direct product of $m$ pure
states $|0\rangle$ and a $(n-m)$-qubit generalized GHZ state
\begin{eqnarray}\label{nqubit}
|\mathcal{G}^p_{2}\rangle_{qubit}  = \cos \phi |0^{[n]} \rangle +
\sin \phi |0^{[m]}1^{[n-m]} \rangle,
\end{eqnarray}
in which $C^{(2)}=(n-m-1)\mathcal{H}_{2}(\phi)$ and $C^{(n-m)}=
\mathcal{H}_2 (\phi) $, but $C^{(n)}=0$.

\subsection{MS states}

 The set of MS states \cite{MS} for qubit case is
an important example to investigate the fundamental concepts in
multipartite system \cite{MSPRL}. We generalize the definition of MS
states to $n$-qutrit system
\begin{eqnarray}\label{MS}
|S\rangle=\frac{1}{
\sqrt{3}}(|\b{0}\rangle+|\b{1}\rangle+|\b{2}\rangle),
\end{eqnarray}
where $|\b{i}\rangle=[\cos \alpha + \sin\alpha(X_1 +
X_1^{\dag})/\sqrt{2}]|i^{[n]}\rangle$ with the operator of $j$-th
qutrit $X_j= |0\rangle \langle1|+ |1\rangle \langle2| +|2\rangle
\langle0|$ and $\alpha \in (0,\arctan\sqrt{2}]$. The operators $X_j$
satisfy $X_j^2=X_j^{\dag}$, ${X^{\dag}_j}^2=X_j$ and $X_j
X^{\dag}_j= X^{\dag}_j X_j=1$.

\emph{Theorem 3.} In the $n$-qutrit MS state (\ref{MS}), only
irreducible $2$-party and $(n-1)$-party correlations exist as
\begin{eqnarray}\label{CMS}
C^{(2)}&=&\mathcal{H}_3(\chi,\pi/4)+(n-2)\ln 3,\nonumber \\
C^{(n-1)}&=&\ln3,
\end{eqnarray}
where $\chi$ satisfies $\cos^2 \chi =\frac{1}{6}(3- \cos 2\alpha +2
\sqrt{2} \sin 2\alpha)$, which is determined by the eigenvalue of
the reduced density matrix $\rho^{(1)}_s$ for the first qutrit.

\begin{proof}
The operator $Q=\sum^{n}_{j=3}Q_{2j}$ has $9$ eigenvectors as
$|i\rangle\otimes |j^{[n-1]}\rangle$ with $i,j=0,1,2$, corresponding
to its maximal eigenvalues $q_{max}=n-2$. We construct an operator
$X=\cos \beta Q_{12}+ \frac{1}{2}\sin \beta (X_1
+X^{\dag}_1+\prod^{n}_{j=2}X_j+\prod^{n}_{j=2}X^{\dag}_j)$ commuting
with $Q$. Thus, the eigenvector of $Q+X$ with the maximal eigenvalue
is the one of $X$ in the subspace $\{|i\rangle\otimes
|j^{[n-1]}\rangle \}$. Choosing the value of $\beta$ satisfying
$\cot \beta = \sqrt{2} (\cot \alpha -\tan \alpha)+1$, one can check
that the MS state (\ref{MS}) is the unique eigenvector corresponding
to the maximal eigenvalues (UEME) of $Q+X$. Therefore, the MS state
can be viewed as the limit of a state without irreducible $n$-party
correlation
\begin{eqnarray}\label{ExpMS}
\rho_s = |S\rangle \langle S|=\lim_{\gamma \rightarrow + \infty}
\Exp \bigr( \eta + \gamma Q + \gamma X \bigr),
\end{eqnarray}
where $ \eta=- \ln \tr [ \Exp (\gamma Q + \gamma X )]$, which leads
to $\tilde{\rho}_{s,n-1}=\rho_s$.

The two-partite operator $R_{12}=[\cos \alpha + \sin\alpha(X_1 +
X_1^{\dag})/\sqrt{2}]Q_{12}[\cos \alpha + \sin\alpha(X_1 +
X_1^{\dag})/\sqrt{2}]$ shares the similar properties with $Q_{ij}$,
$R_{12}^2=R_{12}$ and $\tr_{1,2} R_{12}=3$. The maximal eigenvalue
of $Q+R_{12}$ is triple degenerate with the eigenvectors
$|\b{i}\rangle$. Consequently, the quantum state with only
irreducible $2$-party correlation
\begin{eqnarray}\label{ExpMS2}
\sigma_{s}(\gamma)=\Exp (\eta+\gamma Q + \gamma R_{12} )
\end{eqnarray}
has the limit $\sigma_{s}|_{\gamma \rightarrow \rightarrow +\infty}=
\mathcal{D}_{s}=\sum^{2}_{i=0} \frac{1}{3} | \b{i}\rangle \langle
\b{i}|$, where the value of $\eta$ is determined by the
normalization condition $\tr \sigma_{s}(\gamma)=1$. Thus, we can
take $\tilde{\rho}_{s,m}=\mathcal{D}_s$ for $m=2,3,...n-2$ and
obtain the results in Eq. (\ref{CMS}).
\end{proof}

According to the results by Feng \emph{et al.} \cite{QIC}, it is
easy to identify the $n$-qutrit MS state $|S\rangle$ can be
determined by its $(n-1)$-partite reduced density matrices among
$n$-qutrit pure states. Our results show there is no irreducible
$n$-qutrit correlation in the state $|S\rangle$, which indicates it
can be determined by its $(n-1)$-partite reduced density matrices
among arbitrary $n$-qutrit states (pure or mixed).

The same scheme can be used to deal with the $n$-qubit MS state
\begin{eqnarray}\label{MSqubit}
|S\rangle_{qubit}=(\cos \alpha + \sin \alpha \sigma^x_1)\frac{1}{
\sqrt{2}}(|0^{[n]}\rangle+|1^{[n]}\rangle),
\end{eqnarray}
where $\sigma^{x,y,z}_j$ denote the Pauli operators for the $j$-th
qubit and $\alpha \in (0,\pi/4]$. Replacing  $X_j$ and $Q_{ij}$ by
$(\sigma^x_j+ i \sigma^y_j )/2$ and $(1+\sigma^z_i \sigma^z_j)/2$,
one can construct the operators corresponding to $Q$, $X$ and
$R_{ij}$ in the states (\ref{ExpMS}) and (\ref{ExpMS2}), and obtain
the nonzero correlations
\begin{eqnarray}\label{CMSqubit}
C^{(2)}&=&\mathcal{H}_2(\xi)+(n-2)\ln 2,\nonumber \\
C^{(n-1)}&=&\ln2,
\end{eqnarray}
with $\xi=\pi/4 - \alpha$. The results indicate our generalization
of the MS state to the $n$-qutrit system in Eq. (\ref{MS}) has the
similar correlation distribution with the qubit case.

\section{Conclusion and Discussion}


Following the idea of Zhou's continuity approach, for an $n$-qutrit
quantum state
 $\rho^{[n]}$ without maximal rank, we construct
 $\tilde{\rho}^{[n]}_{m}$ as the limit of a series $n$-qutrit states in the standard exponential form (\ref{exp}). In this way, we
 obtain the degrees of irreducible  multiparty correlations in three families
 $n$-qutrit states, which can be viewed as three generalizations of
 the original GHZ state. The distribution of the total correlations
 in the generalized GHZ state (\ref{G1p}) is the same as the
 $n$-qubit case.  Whereas, in the $n$-qutrit pure state (\ref{G2p}),
 there exist three nonzero irreducible correlations which are
 $C^{(2)},C^{(n)}$ and $C^{(n-m)}$($0<m<n-2$). By contrast, only $C^{(2)}$ and $C^{(n)}$ are nonzero in the $n$-qubit
 generalize GHZ state which is the only pure $n$-qubit state with irreducible
 $n$-party correlation.
This indicates the classification of the total
 correlations for a multipartite pure state in high-dimensional
 system would be more complicated. Also,
an interesting question is raised, which kind of the irreducible
multiparty correlations can simultaneously be nonzero in a pure
states.


 To prove the absence of irreducible $n$-qutrit correlation in the MS state (\ref{MS}),
 we construct an operator $Q+X$ which is a sum of $(n-1)$-partite
 operators and with $|S\rangle$ being its UEMS. The conclusion is proved
 by Eq. (\ref{ExpMS}) in the continuity approach. This indicates the
 open problem about the GSD states is equivalent to the following one
 in the sense of limit: Whether the GSD state is precisely the pure
 states which can't be viewed as the UEME of an operator $Q^{[n-1]}$, a sum of
 $(n-1)$-partite operators. It is straightforward to prove the
 sufficiency part that, there exists no $Q^{[n-1]}$ whose UEME being a GSD state. In the results of
 \cite{QIC}, for an $n$-partite GSD state $|\psi \rangle$, there exist two projectors $P_{1}=\bigotimes^{n}_{j=1}P^{(1)}_{j}
 $ and $P_{2}=\bigotimes^{n}_{j=1}P^{(2)}_{j}$, such that $|\psi \rangle=P_{1} |\psi \rangle + P_{2} |\psi \rangle
 $. The projectors $P^{(1)}_{j}$ and $P^{(2)}_{j}$ for the $j$-th partite
 satisfy $P^{(1)}_{j}|\psi \rangle \neq 0$, $P^{(2)}_{j}|\psi \rangle \neq 0$ and $P^{(1)}_{j}\perp P^{(2)}_{j}
 $. For any $Q^{[n-1]}$, $\langle \psi | Q^{[n-1]} | \psi \rangle =\langle \psi' | Q^{[n-1]} | \psi'
 \rangle$, where $|\psi' \rangle=P_{1} |\psi \rangle - P_{2} |\psi \rangle
 $. Then, the necessity part of this question is left as, how to
 construct a sum of $(n-1)$-partite operators, $Q^{[n-1]}$, whose UEMS is the given non-GSD
 state. We hope to find an universal generalization of the
 construction in Theorem 3 to an arbitrary  non-GSD
 state in our subsequent investigation.

\begin{acknowledgments}
FLZ thanks D. L. Zhou, Y. Feng and Z. H. Ma for their
correspondences and discussions. FLZ is supported by NSF of China
(Grant No. 11105097). JLC is supported by National Basic Research
Program (973 Program) of China under Grant No. 2012CB921900, NSF of
China (Grant Nos. 10975075 and 11175089) and also partly supported
by National Research Foundation and Ministry of Education, Singapore
(Grant No. WBS: R-710-000-008-271).
\end{acknowledgments}

\bibliography{QutritCorrelation}

\end{document}